\documentclass[journal]{IEEEtran}
\ifCLASSINFOpdf
\else
\fi
\usepackage{cite}
\usepackage{amsmath,amssymb,amsfonts,mathrsfs,bm,pifont}
\usepackage{newtxmath}

\usepackage{enumitem}
\usepackage{algorithmic}
\usepackage{graphicx}
\usepackage{textcomp}
\usepackage{xcolor}
\usepackage{subfigure}

\usepackage{hyperref}

\usepackage{algorithmic}

\newcommand{\vj}{{\boldsymbol j}}

\newcommand{\mA}{{\bm A}}
\newcommand{\mB}{{\bm B}}
\newcommand{\mD}{{\bm D}}
\newcommand{\mE}{{\bm E}}
\newcommand{\mF}{{\bm F}}
\newcommand{\mG}{{\bm G}}
\newcommand{\mH}{{\bm H}}
\newcommand{\mI}{{\bm I}}
\newcommand{\mU}{{\bm U}}
\newcommand{\mV}{{\bm V}}
\newcommand{\mP}{{\bm P}}
\newcommand{\mQ}{{\bm Q}}
\newcommand{\mX}{{\bm X}}
\newcommand{\mY}{{\bm Y}}
\newcommand{\mZ}{{\bm Z}}
\newcommand{\mR}{{\bm R}}
\newcommand{\mS}{{\bm S}}
\newcommand{\mT}{{\bm T}}
\newcommand{\mW}{{\bm W}}

\newcommand{\vzero}{{\bm 0}}

\usepackage{booktabs}
\usepackage{multirow}

\begin{document}
\title{Fast Nonconvex $T_2^*$ Mapping Using ADMM}
\author{Shuai Huang, James J. Lah, Jason W. Allen and Deqiang Qiu
\thanks{This work is suppported by National Institutes of Health under Grants R21AG064405 and P50AG025688. \emph{(Corresponding author: Deqiang Qiu.)} }
\thanks{S. Huang, J. W. Allen and D. Qiu are with the Department of Radiology and Imaging Sciences, J. J. Lah is with the Department of Neurology, Emory University, Atlanta, GA 30322 (e-mail: shuai.huang@emory.edu; jlah@emory.edu; jason.w.allen@emory.edu; deqiang.qiu@emory.edu).}}


\maketitle

\begin{abstract}
Magnetic resonance (MR)-$T_2^*$ mapping is widely used to study hemorrhage, calcification and iron deposition in various clinical applications, it provides a direct and precise mapping of desired contrast in the tissue. However, the long acquisition time required by conventional 3D high-resolution $T_2^*$ mapping method causes discomfort to patients and introduces motion artifacts to reconstructed images, which limits its wider applicability. In this paper we address this issue by performing $T_2^*$ mapping from undersampled data using compressive sensing (CS). We formulate the reconstruction as a nonconvex problem that can be decomposed into two subproblems. They can be solved either separately via the standard approach or jointly via the alternating direction method of multipliers (ADMM). Compared to previous CS-based approaches that only apply sparse regularization on the spin density $\mX_0$ and the relaxation rate $\mR_2^*$, our formulation enforces additional sparse priors on the $T_2^*$-weighted images at multiple echoes to improve the reconstruction performance. 
%To avoid the exhaustive search for a proper scaling factor of $\mR_2^*$, we further derive an approximated linear model to the monoexponential decay model of signal magnitude, and use it to compute the least square fit of $\mX_0$ and $\mR_2^*$. 
We performed convergence analysis of the proposed algorithm, evaluated its performance on \emph{in vivo} data, and studied the effects of different sampling schemes. Experimental results showed that the proposed joint-recovery approach generally outperforms the state-of-the-art method, especially in the low-sampling rate regime, making it a preferred choice to perform fast 3D $T_2^*$ mapping in practice. The framework adopted in this work can be easily extended to other problems arising from MR or other imaging modalities with non-linearly coupled variables. %where the variables to be estimated are non-linearly coupled due to physics.
\end{abstract}


% \fi
\IEEEpeerreviewmaketitle

\section{Introduction}
\label{sec:introduction}
Conventional magnetic resonance (MR) imaging techniques such as $T_1$-weighted, $T_2$-weighted imaging are not quantitative, and the produced images cannot be directly compared across different acquisition protocols or scanners. There has since been increasing interest in developing quantitative MR methods that directly measure properties such as spin density, longitudinal and transverse relaxation rates. %Quantitative imaging could improve the sensitivity to lesion as well as the comparability of images acquired from different scanners. 
Here we focus on studying the $T_2^*$ relaxation process that is useful in a number of clinical applications\cite{Fazekas:1999,Kinoshita:2000,ORegan:2009,Yamada:1996,Gupta:2001,Anderson:2001,McNeill:2008} . Apart from the typical spin-spin relaxation (i.e. $T_2$ relaxation), $T_2^*$ relaxation also takes into account the transverse magnetization decay caused by magnetic field inhomogeneity. As a result, $T_2^*$ is shorter than $T_2$, and they are connected by the following relationship \cite{MRI:Stark:2006}
\begin{align}
\label{eq:t2_star_t2}
    \frac{1}{T_2^*}=\frac{1}{T_2}+\frac{1}{{T_2}_{\textrm{in}}}\,,
\end{align}
where $1/{{T_2}_{\textrm{in}}}=\gamma\Delta B_{\textrm{in}}$, $\gamma$ is the gyromagnetic ratio, $\Delta B_{\textrm{in}}$ is the magnetic filed inhomogeneity caused by magnetic susceptibility differences within the tissues, chemical shift, and gradient fields applied for spatial encoding \cite{MRI:Nishimura:2010}. %Note that the inhomogeneity $\Delta B_{\textrm{in}}$ can be eliminated in a spin-echo (SE) sequence by applying a $180^\circ$ pulse. 
In order to preserve the inhomogeneity $\Delta B_{\textrm{in}}$, we use the gradient-echo (GRE) sequence to acquire the data. No $180^\circ$ refocusing pulse is applied after the radio-frequency (RF) excitation, and measurements are acquired at multiple echo times (TE) within one repetition time (TR).

The tissue contrasts revealed by $T_2^*$-weighted images have proved quite useful in studying hemorrhage \cite{Fazekas:1999,Kinoshita:2000,ORegan:2009}, calcification \cite{Yamada:1996,Gupta:2001}, and iron deposition \cite{Anderson:2001,McNeill:2008} in various tissues and diseases. However, the \emph{qualitative} contrast information obtained from a series of $T_2^*$-weighted images depends on the sampling pulse sequence and imaging platform. Furthermore, qualitative $T_2^*$ image interpretation is subjective and may vary between readers. As a result, \emph{quantitative} measurement of the $T_2^*$ relaxation values has attracted a lot of interests in recent years, since it provides a direct and precise mapping of desired contrast in the tissue \cite{Chavhan:2009:T2StarReview}. $T_2^*$ mapping has been used in a variety of clinical applications such as susceptibility-weighted imaging \cite{Haacke:SWI:2009,Mittal:SWI:2009}, perfusion MR imaging \cite{Soonmee:PMRI:2002,Weber:PMRI:2006}, functional MR imaging \cite{Gore:FMRI:2003,Poldrack:FMRI:2011} and iron overload imaging \cite{Hankins:Iron:2009,Tucci:Iron:2008}. In particular, the quantitative nature of $T_2^*$ mapping has made it possible to detect and monitor small pathological changes in Alzheimer's disease (AD) \cite{Aquino:AD:2009,Callaghan:AD:2014}. An animal study \cite{OCallaghan:TissueMS:2017} has shown that $T_2^*$ relaxation values in the tissue are sensitive to early changes in Tau pathology, a key neuro-pathological hallmark of AD. This could provide valuable biomarker information in helping the clinicians to predict risk of transition from normal cognition to mild cognitive impairment (MCI), and the rate of deterioration in MCI patients.

Three-dimensional (3D) high-resolution imaging is usually needed to find early predictive biomarkers of diseases such as AD, so that early pathological changes could be localized within small anatomical structures \cite{Henneman:VMRI:2009}. High-resolution $T_2^*$ mapping has also been found to be useful in studying iron loading in Parkinson's Disease \cite{Langkammer:2010:BrainIron,Wang:2016:BrainIron}, as well as in the quantification of the liver overloading \cite{Sirlin:2010:LiverIron,Hernando:2014:LiverIron}. The long acquisition time required by conventional $T_2^*$ mapping methods causes discomfort to the patients, introduces motion artifacts to the reconstructed images, and reduces the throughput of clinical imaging studies. Developing fast imaging methods that operate at lower sampling rates to reduce the scan time is thus of great importance in practice. In this paper we use compressive sensing (CS) techniques \cite{CS06,Candes:CS:2008} to perform $T_2^*$ mapping from undersampled data. We formulate the reconstruction into the following two subproblems: 
\begin{enumerate}
    \item Reconstruction of the magnitude images $\mX_i$ and the phase images $\mZ_i$ at multiple echo times $i\in\{1,\cdots,E\}$.
    \item Reconstruction of the spin density $\mX_0$ and the relaxation rate $\mR_2^*=\frac{1}{T_2^*}$.
\end{enumerate}
They can be solved either separately via the standard approach or jointly via the alternating direction method of multipliers (ADMM) with convergence guarantee \cite{Boyd:ADMM:2011,Wang:ADMMNonconvex:2015}. The images $\mX_0,\mX_i,\mR_2^*$ can be considered to be approximately sparse in some proper basis like the wavelet basis \cite{DBWav92}. This sparse prior information can be leveraged to improve reconstruction performance by adding regularization to promote sparse solutions. Compared to previous CS-based methods that can only apply regularization on $\mX_0$ and $\mR_2^*$, our ADMM formulation allows us to make use of the sparse priors on multi-echo images $\mX_i$ and apply regularization on them as well. Experimental results show that the proposed joint recovery approach generally outperforms the state-of-the-art model-based approach, especially in the low-sampling rate regime.

We undersampled the k-space data using the Poisson disk sampling scheme \cite{Dunbar:PoissonDisk:2006,Bridson:PoissonDisk:2007}, which imposes a minimum distance $d_{\min}$ between any two sampling locations. Note that the regular random sampling scheme is a special case of the Poisson disk sampling scheme when $d_{\min}$ is set to $0$. With a proper $d_{\min}$, Poisson disk sampling could achieve a much more uniform sampling distribution than the regular random sampling. In an effort to acquire incoherent measurements, we also investigated the use of sampling patterns that are complementary at different echo times to increase the overall coverage of k-space across multiple echoes. We conducted experiments to study how this could affect the performance in $T_2^*$ mapping, and find that using complementary sampling patterns performs slightly better than using identical sampling pattern across different echoes. In practice we can keep the sampling pattern fixed across different echo times to simplify the pulse sequence programming on the MRI scanner.

\subsection{Prior Work}
Parallel imaging coupled with multi-channel phase-array coils is a widely adopted method that explores the redundancy in receiver coils to allow undersampling in the k-space \cite{Deshmane:ParallelMR:2012,Pruessmann:SENSE:1999,Griswold:GRAPPA:2002}. It is a generic method in that it is not designed for a specific clinical task, but can be used in various MRI applications. In this paper we use multi-channel phase-array coils to improve the signal-to-noise ratio (SNR) of the reconstruction. Both the $T_2^*$ and $T_2$ relaxations can be modeled by a monoexponential decay of the transverse magnetization. As a result, the methods developed for $T_2^*$ and $T_2$ mappings can be used interchangeably. In the following we shall review previous works on reconstructing $T_2^*$ and $T_2$ mappings from undersampled data together. 

Various approaches have been proposed to perform $T_2$ mapping. For example, echo sharing is used to reconstruct $T_2$-weighted images from a single radial fast spin echo (SE) dataset \cite{Song:KWIC:2000,Altbach:ES:2005}. The multi-echo images are first reconstructed by mixing different data acquired at speciﬁc echo times in the central k-space region with the same radial data in the outer k-space region. The $T_2$ map can then be obtained via pixel-wise monoexponential fits of the $T_2$-weighted images. However, the mixing of high frequency TE data introduces errors to the high frequency details in the $T_2$ map. To avoid mixing different echo data, model-based iterative approaches can be used to directly reconstruct the spin density $\mX_0$ and the relaxation rate $\mR_2$ from the radial fast SE data \cite{Block:ModelT2:2009,Sumpf:Model:2011}. A nonlinear inverse approach that minimizes the regularized least squared error using conjugate gradient (CG) was proposed in \cite{Block:ModelT2:2009}. Due to the nonlinearity of the exponential decay, the CG method is sensitive to the scaling of $\mR_2$. Its speed and success heavily depend on choosing a suitable scaling factor. A data-driven method can be used to estimate the scaling factor based on low resolution reconstructions from central k-space data \cite{Sumpf:Model:2011}. Alternatively, the nonlinear exponential decay can be approximated by a linear combination of several pre-computed principal components, thus eliminating the need for scaling \cite{Huang:T2mapping:2012}.

As shown in \eqref{eq:t2_star_t2}, $T_2^*$ is shorter than $T_2$ due to magnetic field inhomogeneity. The SNRs of the measurements thus decrease faster, creating a signal floor offset at longer echo times in the GRE sequence. The monoexponential decay model no longer accurately describes the signal decay across multiple echoes. A constant $C$ was added to the monoexponential model $\mX_i=\mX_0\cdot\exp(-t_i\cdot\mR_2^*)+C$ in \cite{Ghugre:R2Star:2006} to correct the signal floor offset. Another method is to truncate the measurements that do not fit into the monoexponential model at longer echo times \cite{He:T2StarTrunc:2008,He:T2StarTrunc:2013}. Both methods have their pros and cons in estimating $\mR_2^*$. Empirically, the offset method tends to produce higher $R2^*$ values than the truncation method \cite{Triadyaksa:2019}. In this paper we use the truncation methods when performing $T_2^*$ mapping.

\subsection{Our Contribution and Paper Outline}
We propose to formulate the reconstruction of $T_2^*$ mapping from undersampled data into two subproblems: one that recovers the $T_2^*$-weighted images $\mX_i$ at each echo time, and one that recovers the spin density $\mX_0$ and the relaxation rate $\mR_2^*$. The two subproblems can be solved either separately via the standard approach or jointly via the ADMM approach with convergence guarantee. Compared to previous approaches that only enforce sparse priors on $\mX_0$ and $\mR_2^*$, our formulation allows us to make use of additional sparse priors on $\mX_i$ to obtain better reconstructions. To avoid the exhaustive search for a proper scaling factor of $\mR_2^*$ due to the nonlinearity of the monoexponential decay model, we derive an approximated linear model to compute the regularized least square fit of $\mX_0$ and $\mR_2^*$. Experimental results show that the proposed approach outperforms the state-of-the-art model-based approach, especially in the low-sampling rate regime.

This paper proceeds as follows. In Section \ref{sec:problem_formulation} we present the problem formulation of the $T_2^*$ mapping, and introduce the proposed decoupled and joint recovery approaches. In Section \ref{sec:propose_admm} we solve the joint recovery problem using ADMM, show that the simpler decoupled problem is equivalent to one ADMM iteration of the joint recovery problem, and perform the convergence analysis. In Section \ref{sec:exp} we compare the proposed approaches with the state-of-the-art model-based approach by performing image reconstruction experiments from \emph{in vivo} data under varying sampling rates. We also compare the effects of different k-space undersampling schemes. We finally conclude this paper with a discussion in Section \ref{sec:conclusion}.

\section{Problem Formulation}
\label{sec:problem_formulation}
\begin{figure}[tbp]
\centering
\vspace{-1em}
\subfigure[]{
\includegraphics[height=.2\textwidth]{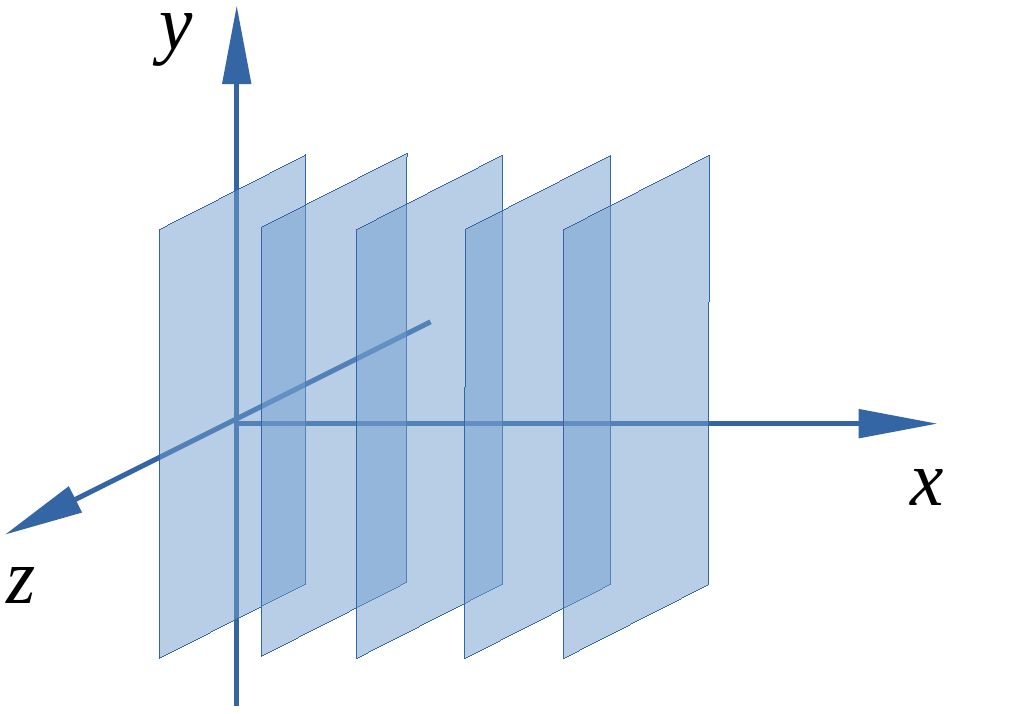}}
\subfigure[]{
\includegraphics[height=.2\textwidth]{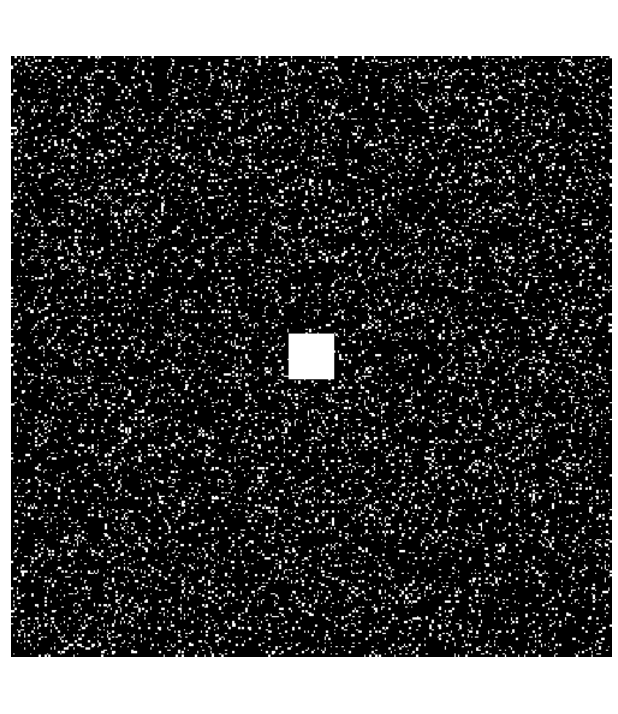}}
\caption{(a) Undersampling in the k-space takes place in the plane determined by the two phase encoding directions $y$ and $z$, whereas the readout direction $x$ is fully sampled; (b) The Poisson disk sampling pattern in the $y$-$z$ plane, the central k-space is fully sampled and used to estimate the sensitivity maps of the receiver coils. } 
\vspace{-1em}
\label{fig:k_space_sampling}
\end{figure}

In this paper we aim to develop a fast 3D imaging method that allows us to reconstruct $T_2^*$ maps along with spin (proton) density maps and phase images from undersampled k-space data. As shown in Fig. \ref{fig:k_space_sampling}, undersampling takes place in the plane of the two phase encoding directions $y$ and $z$, whereas the readout direction $x$ is fully sampled. The Poisson disk sampling scheme is used to select the sampling locations in the $y$-$z$ plane, it imposes a minimum pairwise distance constraint between any two sampling locations, producing a more uniform sampling distribution than random sampling. By performing 1D FFT on the 3D k-space data along the readout direction, we can get 2D k-space data from different 2D slices. This further enables us to reconstruct the maps from multiple slices in parallel to save time. We thus focus on 2D reconstructions in the following discussion, which can be easily extended to the 3D case in a straightforward manner.

\begin{figure}[tbp]
\centering
\vspace{-1em}
\includegraphics[width=.4\textwidth]{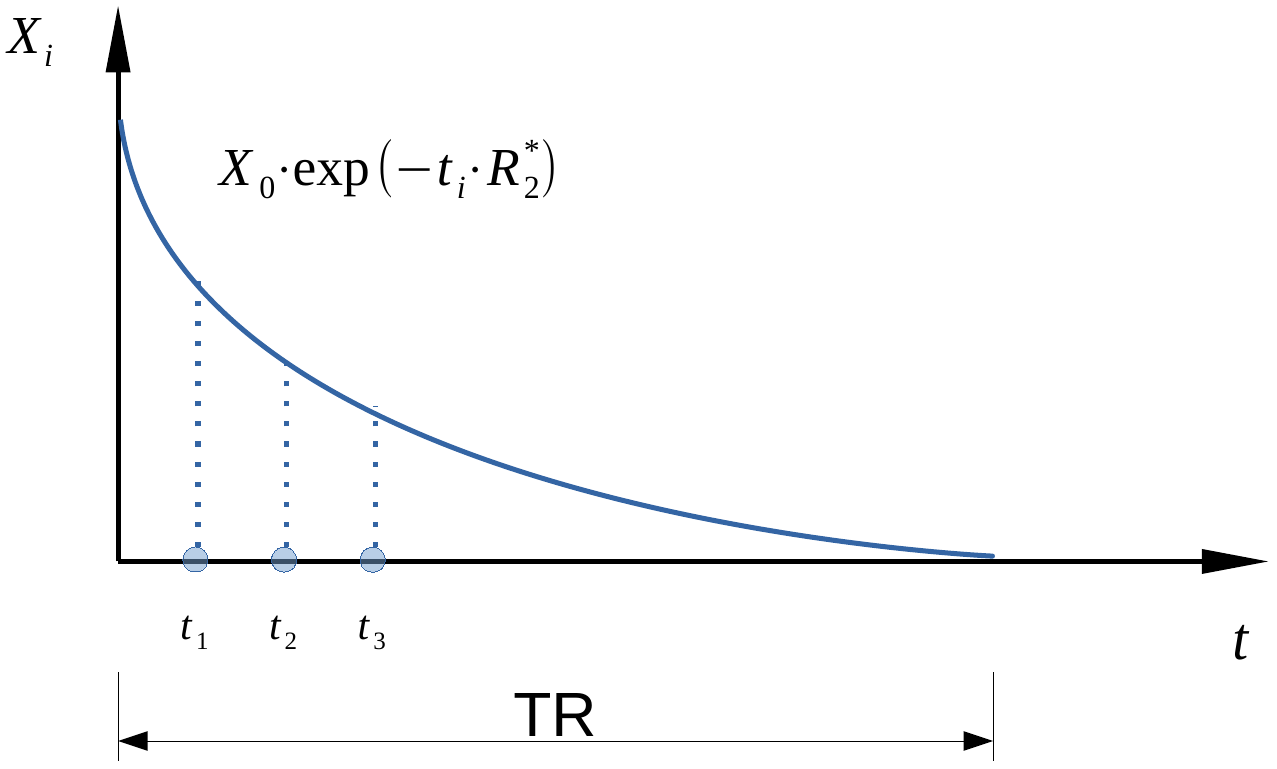}
\caption{The signal magnitude follows the monoexponential decay model. For the $T_2^*$ mapping, the k-space data is sampled at multiple echo times (TE) $t_i$ within one repetition time (TR) of a gradient-echo (GRE) sequence.}
\vspace{-1em}
\label{fig:monoexponential_te}
\end{figure}

As shown in Fig. \ref{fig:monoexponential_te}, the k-space data is sampled at multiple echo times (TE) within one repetition time (TR) of a gradient-echo sequence (GRE). The magnetization at every voxel across different echo times can be modeled by the exponential decay \cite{MRI:Nishimura:2010}
\begin{align}
\label{eq:exp_decay}
    \mX_i=\mX_0\cdot\exp\left(-t_i\cdot\mR_2^*\right)\,,
\end{align}
where $\mX_i$ is the $i$-th echo image at the time $t_i$, $\mX_0$ is the spin density image, and $\mR_2^*=\frac{1}{T_2^*}$ is the relaxation rate image. Multiple receiver coils are used to acquire the measurements to improve the overall SNR. Let $\mY_{ij}$ denote the measurements from the $j$-th receiver coil at time $t_i$. We have
\begin{align}
\label{eq:measurement_model}
\begin{split}
    %\mY_{ij}&=\mA_i\mS_j\mZ_i\mX_i\\
    \mY_{ij}&=\mA_i\mS_j\mZ_i\mX_0\cdot\exp\left(-t_i\cdot\mR_2^*\right)\,,
\end{split}
\end{align}
where $\mA_i=\mP_i\mF$ is the sampling operator at the time $t_i$, with $\mP_i$ and $\mF$ being the undersampling matrix and Fourier operator respectively, $\mS_j$ is the sensitivity map of the $j$-th receiver coil, $\mZ_i$ is the phase image at the time $t_i$. $\mS_j,\mZ_i,\mX_0,\mR_2^*$ are all functions of locations and are represented as discrete images. The sensitivity map $\mS_j$ can be estimated using the ESPIRiT approach \cite{Uecker:ESPIRiT:2014}. Here we would like to recover $\mR_2^*$ along with $\mX_0,\mZ_i$ from the undersampled measurements $\mY_{ij}$.

\subsection{$T_2^*$ Mapping via Multi-echo Regularization}
Since both the $T_2$ and $T_2^*$ relaxations of the transverse magnetization can be described in \eqref{eq:exp_decay}, the methods developed for $T_2$ and $T_2^*$ mappings can be used interchangeably. Conjugate gradient has been used to perform the least square fit of the monoexponential decay model \cite{Block:ModelT2:2009}. However, due to the nonlinearity of the model, proper scaling between $\mX_0$ and $\mR_2$ is needed to achieve convergence within a reasonable time, and the choice of the scaling parameter is data-dependent and sensitive \cite{Sumpf:Model:2011}. By taking the ``$\log$'' of both sides of \eqref{eq:exp_decay}, we can establish a linear relationship between $\log\mX_0$ and $\mR_2^*$
\begin{align}
\label{eq:log_exp_decay}
    \log\mX_i=\log\mX_0-t_i\cdot\mR_2^*\,.
\end{align}
Although the least square fit of the above \eqref{eq:log_exp_decay} can be obtained easily, it puts higher weights on the measurements acquired at later (larger) echo times that have lower SNR \cite{Fraile:2005}. In this paper we shall perform a \emph{weighted} least square fit of \eqref{eq:log_exp_decay} and minimize the following loss function
\begin{align}
\label{eq:weighted_log_exp_lsq_fit}
    \sum_ix_i^2\cdot\left(\log x_0-t_i\cdot r_2^*-\log x_i\right)^2=\sum_iw_i\cdot e_i\,.
\end{align}
A weight $w_i=x_i^2$ is multiplied to the least square error $e_i$ at time $t_i$:
\begin{align}
\begin{split}
e_i&=\left(\log x_0-t_i\cdot r_2^*-\log x_i\right)^2\\
% &=\log^2\left(\frac{x_0\exp(-t_i\cdot r_2^*)-x_i}{x_i}+1\right)\\
&=\log^2\left(p_i+1\right)\,,
\end{split}
\end{align}
where $p_i=\frac{x_0\exp(-t_i\cdot r_2^*)-x_i}{x_i}$. In fact, if we take the first order approximation of $\log(p_i+1)$ using the Mercator series when $p_i$ is small, we can get the least square fit of the original monoexponential decay model. Specifically, we have
\begin{align}
\label{eq:mercator_series}
\begin{split}
    \log(p_i+1)= p_i-\frac{p_i^2}{2}+\frac{p_i^3}{3}-\frac{p_i^4}{4}+\cdots\,.
\end{split}
\end{align}

Ignoring the terms of second order and beyond, we can get $\log(p_i+1)\approx p_i$ when $p_i$ is small. Plugging it into \eqref{eq:weighted_log_exp_lsq_fit}, we have
\[
    \sum_ix_i^2\cdot\left(\log x_0-t_i\cdot r_2^*-\log x_i\right)^2\approx\sum_i\left(x_0\exp(-t_i\cdot r_2^*)-x_i\right)^2.
\]
We can see that the ``weighted least square fit'' of the linearized model in \eqref{eq:log_exp_decay} can be viewed as an \emph{approximation} to the ``least square fit'' of the monoexponential decay model in \eqref{eq:exp_decay}.

The images $\mX_i,\mX_0,\mR_2^*$ can be considered to be approximately sparse in some proper basis such as the wavelet basis \cite{DBWav92}. Previous methods only enforce the sparse prior information during reconstruction via the regularization on $\mX_0$ and $\mR_2^*$ \cite{Block:ModelT2:2009,Huang:T2mapping:2012,Zhao:PM:2015}. In this paper we shall reformulate the reconstruction problem and introduce additional regularization on the multi-echo images $\mX_i$. This way we could take full advantage of the sparse priors on all the images and achieve better reconstruction performance. 

There are two ways to approach the reconstruction of $\mX_i,\mX_0,\mR_2^*$. The first way is to decouple the problem into the reconstruction of $\mX_i$ and the reconstruction of $\mX_0,\mR_2^*$. The second way is to reconstruct $\mX_i,\mX_0,\mR_2^*$ jointly. Letting $\mA_{ij}=\mA_i\mS_j$ and $\mH_0=\log\mX_0$, we have the following two problems:

\subsubsection{The decoupled recovery}
%\begin{decoupled_recovery}
\begin{align}
    \label{eq:decoupled_xi}
    \min_{\mZ_i,\mX_i}\quad&\sum_{ij}\left\|\mY_{ij}-\mA_{ij}\mZ_i\mX_i\right\|_2^2+\lambda_1\sum_i\left\|\Phi(\mX_i)\right\|_1\\
    \begin{split}
    \label{eq:decoupled_h0_r2}
    \min_{\mH_0,\mR_2}\quad&\sum_i\mX_i^2\|\mH_0-t_i\mR_2^*-\log\mX_i\|_2^2\\
    &+\lambda_2\left\|\Phi(\mH_0)\right\|_1+\lambda_3\left\|\Phi(\mR_2^*)\right\|_1\,,
    \end{split}
\end{align}
where $\sum_{ij}\left\|\mY_{ij}-\mA_{ij}\mZ_i\mX_i\right\|_2^2$ is the data fidelity term, $\Phi(\cdot)$ is the wavelet transform, $\{\lambda_1,\lambda_2,\lambda_3\}$ are the regularization parameters, $\left\{\lambda_1\left\|\Phi(\mX_i)\right\|_1\right.$, $\lambda_2\left\|\Phi(\mH_0)\right\|_1$, $\left.\lambda_3\left\|\Phi(\mR_2^*)\right\|_1\right\}$ are the regularization terms to enforce the sparse priors on $\{\mX_i,\mH_0,\mR_2^*\}$, and $\sum_i\mX_i^2\|\mH_0-t_i\mR_2^*-\log\mX_i\|_2^2$ is the \emph{weighted} least square loss to enforce the monoexponential decay model in \eqref{eq:exp_decay}. In this paper, we use the sparsity averaging method \cite{SACS13} to construct an
over-complete wavelet basis $\Phi(\cdot)$ by concatenating Db1-Db8 wavelets \cite{DBWav92}.
%\end{decoupled_recovery}

\subsubsection{The joint recovery}
%\begin{joint_recovery}
\begin{align}
\label{eq:t2_star_mapping}
\begin{split}
    \min_{\mZ_i,\mX_i,\mH_0,\mR_2^*}\quad&\sum_{ij}\left\|\mY_{ij}-\mA_{ij}\mZ_i\mX_i\right\|_2^2+\lambda_1\sum_i\left\|\Phi(\mX_i)\right\|_1\\
    &+\lambda\sum_i\mX_i^2\|\mH_0-t_i\mR_2^*-\log\mX_i\|_2^2\\
    &+\lambda\cdot\lambda_2\left\|\Phi(\mH_0)\right\|_1+\lambda\cdot\lambda_3\left\|\Phi(\mR_2^*)\right\|_1\,,
\end{split}
\end{align}
where $\lambda$ is the regularization parameter that balances the trade-off between the measurement model in \eqref{eq:measurement_model} and monoexponential decay model in \eqref{eq:exp_decay}.
%\end{joint_recovery}

Later in section \ref{sec:propose_admm} we propose to solve the joint recovery using ADMM. The decoupled recovery can then be viewed as one ADMM iteration in the joint recovery. Although the decoupled recovery is obviously easier to solve, the joint recovery is more robust when there are insufficient measurements. In fact, when the sampling rate is high, the two approaches perform almost equally well. When the sampling rate is low, the joint recovery performs much better than the decoupled recovery.

\section{Proposed Approach}
\label{sec:propose_admm}
Since the decoupled recovery simply amounts to one ADMM iteration in the joint recovery under the right choice of parameters, we shall focus on solving the joint recovery problem using ADMM in this section. For the joint recovery of $\{\mZ_i,\mX_i,\mH_0,\mR_2^*\}$, the nonconvex problem in \eqref{eq:t2_star_mapping} is difficult to solve directly. We can use ADMM \cite{Boyd:ADMM:2011} to decompose it into two easier subproblems, corresponding to the optimization of $\mZ_i,\mX_i$ and the optimization of $\mH_0,\mR_2^*$ respectively. To make it clearer, the objective function can be separated into two parts:
\begin{align}
    \min_{\mZ_i,\mX_i,\mH_0,\mR_2^*}\quad f(\mZ_i,\mX_i)+\lambda\cdot g(\mX_i,\mH_0,\mR_2^*)\,,
\end{align}
where
\begin{align}
    &f(\mZ_i,\mX_i)=\sum_{ij}\left\|\mY_{ij}-\mA_{ij}\mZ_i\mX_i\right\|_2^2+\lambda_1\sum_i\left\|\Phi(\mX_i)\right\|_1\\
    \begin{split}
    &g(\mX_i,\mH_0,\mR_2^*)=\sum_i\mX_i^2\|\mH_0-t_i\mR_2^*-\log\mX_i\|_2^2\\
    %&\phantom{g(\mX_i,\mH_0,\mR_2^*)=}+\lambda_2\left\|\Phi(\mH_0)\right\|_1+\lambda_3\left\|\Phi(\mR_2^*)\right\|_1\,.
    &\quad\quad\quad\quad\quad\quad\quad+\lambda_2\left\|\Phi(\mH_0)\right\|_1+\lambda_3\left\|\Phi(\mR_2^*)\right\|_1\,.
    \end{split}
\end{align}
This is equivalent to the following constrained problem
\begin{align}
\label{eq:t2_star_mapping_separate}
\begin{split}
    \min_{\mZ_i,\mX_i,\mH_0,\mR_2^*,\mE_i}&\quad f(\mZ_i,\mX_i)+\lambda\cdot g(\mE_i,\mH_0,\mR_2^*)\\
    \textrm{subject to}&\quad \mX_i=\mE_i\,.
\end{split}
\end{align}
Although ADMM was originally proposed to solve convex problems, it has been used with success in various nonconvex problems as well. Following the sufficient conditions introduced in \cite{Wang:ADMMNonconvex:2015}, the convergence analysis of the proposed ADMM approach to solve \eqref{eq:t2_star_mapping_separate} is discussed later in section \ref{subsec:convergence_admm}. Specifically, the augmented Lagrangian of the ADMM formulation to perform $T_2^*$ mapping is
\begin{align}
\begin{split}
    \mathcal{L}_\rho= &f(\mZ_i,\mX_i)+\lambda\cdot g(\mE_i,\mH_0,\mR_2^*)\\
    &+\sum_i\mB_i(\mX_i-\mE_i)+\frac{\rho}{2}\|\mX_i-\mE_i\|_2^2\,,
\end{split}
\end{align}
where $\mB_i$ contains the dual variables, and $\rho$ is the regularization parameter that enforces the equality constraint $\mX_i=\mE_i$. The ADMM updates of the solutions are
\begin{enumerate}
    \item Solve for $\mZ_i,\mX_i$:
    \begin{align}
    \label{eq:admm_z_x}
    \begin{split}
        \min_{\mZ_i,\mX_i}\quad&\sum_{ij}\left\|\mY_{ij}-\mA_{ij}\mZ_i\mX_i\right\|_2^2+\lambda_1\sum_i\left\|\Phi(\mX_i)\right\|_1\\
        &+\sum_i\mB_i(\mX_i-\mE_i)+\frac{\rho}{2}\|\mX_i-\mE_i\|_2^2\,.
    \end{split}
    \end{align}
    \item Solve for $\mH_0,\mR_2^*$:
    \begin{align}
    \label{eq:admm_h_r}
    \begin{split}
        \min_{\mH_0,\mR_2^*}\quad&\sum_i\mE_i^2\|\mH_0-t_i\mR_2^*-\log\mE_i\|_2^2\\
        &+\lambda_2\left\|\Phi(\mH_0)\right\|_1+\lambda_3\left\|\Phi(\mR_2^*)\right\|_1\,.
    \end{split}
    \end{align}
    \item Solve for $\mE_i$:
    \begin{align}
    \label{eq:admm_e}
    \begin{split}
        \min_{\mE_i}\quad&\mE_i^2\|\mH_0-t_i\mR_2^*-\log\mE_i\|_2^2\\
        &+\mB_i(\mX_i-\mE_i)+\frac{\rho}{2}\|\mX_i-\mE_i\|_2^2\,.
    \end{split}
    \end{align}
    \item Update $\mB_i$:
    \begin{align}
    \label{eq:admm_b}
        \mB_i = \mB_i+\rho(\mX_i-\mE_i)\,.
    \end{align}
\end{enumerate}
We can see that when $\mB_i=\vzero$, $\rho=0$ and $\mX_i=\mE_i$, the subproblems \eqref{eq:admm_z_x}, \eqref{eq:admm_h_r} then become the decoupled recovery in \eqref{eq:decoupled_xi}, \eqref{eq:decoupled_h0_r2}. The subproblems in \eqref{eq:admm_z_x}-\eqref{eq:admm_e} are easier to solve compared to the original joint recovery problem in \eqref{eq:t2_star_mapping}. We next show how they can be solved one by one.

\subsection{Computation of $Z_i$ and $ X_i$}
We proceed to solve \eqref{eq:admm_z_x} using the proximal gradient method iteratively until convergence \cite{Beck:FISTA:2009,Parikh:Proximal:2013}. Let $\mU_i=\mZ_i\mX_i$ and $f_j(\mU_i)=\|\mY_{ij}-\mA_{ij}\mU_i\|_2^2$. In the $(k+1)$-th iteration, we perform the proximal regularization on it with respect to the solution $\mU_i^{(k)}$ from the previous $k$-th iteration.
%\begin{align}
\[
\begin{split}
    f_j(\mU_i)\leq& f_j(\mU_i^{(k)})+\langle\mU_i-\mU_i^{(k)},\nabla f_j(\mU_i^{(k)})\rangle\\
    &+ \langle\overline{\mU}_i-\overline{\mU}_i^{(k)},\nabla f_j(\overline{\mU}_i^{(k)})\rangle + \frac{\kappa_{ij}}{2}\|\mU_i-\mU_i^{(k)}\|_2^2\\
    =&\frac{\kappa_{ij}}{2}\|\mU_i-\mQ_{ij}^{(k)}\|_2^2+\mathcal{O}(\mU_i^{(k)})\,,
\end{split}
\]
%\end{align}
where $\kappa_{ij}$ is the Lipschitz constant of $f_j(\mU_i)$ to ensure the above proximal regularization holds, $\overline{\mU}_i$ is the conjugate of $\mU_i$, $\mathcal{O}(\mU_i^{(k)})$ is some constant that depends on the previous solution $\mU_i^{(k)}$, the gradient $\nabla f_j(\mU_i)$ and $\mQ_{ij}^{(k)}$ are as follows
\begin{align}
    \nabla f_j(\mU_i)&=\left(\frac{\partial f_j(\mU_i)}{\partial \mU_i}\right)^*=\mA_{ij}^*\mA_{ij}\mU_i-\mA_{ij}^*\mY_{ij}\\
    \mQ_{ij}^{(k)}&=\mU_i^{(k)}-\frac{2}{\kappa_{ij}}\nabla f_j(\mU_i^{(k)})\,.
\end{align}

Following the FISTA approach \cite{Beck:FISTA:2009}, we then solve the following problem in the $(k+1)$-th iteration
\begin{align}
\label{eq:kp1_opt}
\begin{split}
    \min_{\mZ_i,\mX_i}\quad&\sum_{ij}\frac{\kappa_{ij}}{2}\|\mZ_i\mX_i-\mQ_{ij}^{(k)}\|_2^2+\lambda_1\sum_i\left\|\Phi(\mX_i)\right\|_1\\
    &+\sum_i\mB_i(\mX_i-\mE_i)+\frac{\rho}{2}\|\mX_i-\mE_i\|_2^2\,.
\end{split}
\end{align}

\subsubsection{Computation of $Z_i$}
 Let $\mZ_i=\exp(\vj\boldsymbol{\Theta}_i)$, where $\boldsymbol\Theta_i\in[0,2\pi)$ is the angle of the phase image. We then have
    \begin{align}
    \label{eq:opt_theta}
        \min_{\boldsymbol\Theta_i}\sum_j\frac{\kappa_{ij}}{2}\|\mX_i\exp(\vj\boldsymbol\Theta_i)-\mQ_{ij}^{(k)}\|_2^2\,.
    \end{align}
    % Taking the first order derivative of \eqref{eq:opt_theta}, we have
    % \begin{align}
    % \label{eq:der_theta}
    %     \frac{\partial }{\partial \boldsymbol\Theta_i}=\mX_i\sum_j\kappa_{ij}\left(\textnormal{Re}(\mQ_{ij}^{(k)})\sin\boldsymbol\Theta_i-\textnormal{Im}(\mQ_{ij}^{(k)})\cos\boldsymbol\Theta_i\right)\,,
    % \end{align}
    % where $\textrm{Re}(\cdot)$ takes the real coefficient of a complex number, and $\textrm{Im}(\cdot)$ takes the imaginary coefficient of a complex number. Setting the first order derivative \eqref{eq:der_theta} to zero, we have
    Setting the first order derivative of \eqref{eq:opt_theta} to zero, we have
    \begin{align}
    \label{eq:theta_1st_der_zero}
        \frac{\sin\boldsymbol\Theta_i}{\cos\boldsymbol\Theta_i}=\frac{\sum_j\kappa_{ij}\textnormal{Im}(\mQ_{ij}^{(k)})}{\sum_j\kappa_{ij}\textnormal{Re}(\mQ_{ij}^{(k)})}\,.
    \end{align}
    There are two solutions that satisfy \eqref{eq:theta_1st_der_zero}. In order to find the minimizing solution, we further compute the second order derivative of \eqref{eq:opt_theta}
    \begin{align}
    \label{eq:2nd_der_theta}
        \frac{\partial^2}{\partial \boldsymbol\Theta_i^2}=\mX_i\left(\cos\boldsymbol\Theta_i\textstyle\sum_j\kappa_{ij}\textnormal{Re}(\mQ_{ij}^{(k)})+\sin\boldsymbol\Theta_i\textstyle\sum_j\kappa_{ij}\textnormal{Im}(\mQ_{ij}^{(k)})\right)\,.
    \end{align}
    The above \eqref{eq:2nd_der_theta} is positive if $\cos\boldsymbol\Theta_i$ has the same sign as $\sum_j\kappa_{ij}\textnormal{Re}(\mQ_{ij}^{(k)})$ and $\sin\boldsymbol\Theta_i$ has the same sign as $\sum_j\textnormal{Im}(\mQ_{ij}^{(k)})$. Hence the minimizing $\boldsymbol\Theta_i$ is 
    \begin{align}
        \boldsymbol\Theta_i = \textnormal{Ang}\left(\textstyle\sum_j\kappa_{ij}\textnormal{Re}(\mQ_{ij}^{(k)}) + \vj\textstyle\sum_j\kappa_{ij}\textnormal{Im}(\mQ_{ij}^{(k)})\right)\,,
    \end{align}
    where $\textnormal{Ang}(\cdot)$ computes the angle of a complex number.
    
\subsubsection{Computation of $X_i$}
We further apply the proximal regularization of on $\mathscr{h}_j(\mX_i)=\|\mZ_i\mX_i-\mQ_{ij}^{(k)}\|_2^2$ in \eqref{eq:kp1_opt}.
\begin{align}
\begin{split}
    % \mathscr{h}_j(\mX_i)\leq& \mathscr{h}_j(\mX_i^{(k)}) + \langle\mX_i-\mX_i^{(k)},\nabla \mathscr{h}_j(\mX_i^{(k)})\rangle\\
    % &+\frac{\alpha_i}{2}\|\mX_i-\mX_i^{(k)}\|_2^2\\
    % =&\frac{\alpha_i}{2}\|\mX_i-\mP_{ij}^{(k)}\|_2^2+\mathcal{O}(\mX_i^{(k)})\,,
    \mathscr{h}_j(\mX_i)\leq\frac{\alpha_i}{2}\|\mX_i-\mP_{ij}^{(k)}\|_2^2+\mathcal{O}(\mX_i^{(k)})\,,
\end{split}
\end{align}
where $\alpha_i$ is the Lipschitz constant of $\mathscr{h}_j(\mX_i)$ to ensure the proximal regularization holds, $\mathcal{O}(\mX_i^{(k)})$ is some constant that depends on the previous solution $\mX_i^{(k)}$, and $\mP_{ij}^{(k)}$ is
%Since $\mZ_i$ is the phase image, we can choose $\alpha_i=2$. The term $\mP_{ij}^{(k)}$ is as follows
\begin{align}
% \begin{split}
%     %\nabla g_{ij}(\mX_0) &= 2\mX_i\overline{\mZ}_i\mZ_i-(\overline{\mZ}_i\mQ_{ij}^{(k)}+\mZ_i\overline{\mQ}_{ij}^{(k)})\\
%     \nabla \mathscr{h}_j(\mX_i) &=2\mX_i-2\textnormal{Re}(\overline{\mZ}_i\mQ_{ij}^{(k)})
% \end{split}\\
\begin{split}
\mP_{ij}^{(k)} &= \mX_i^{(k)}-\frac{1}{\alpha_i}\nabla \mathscr{h}_j(\mX_i^{(k)})= \textnormal{Re}(\overline{\mZ}_i\mQ_{ij}^{(k)})\,.
\end{split}
\end{align}
We then solve the following $l_1$-minimization problem
% \begin{align}
% \label{eq:kp1_opt_xi}
% \begin{split}
%     \min_{\mX_i}\quad&\sum_{ij}\frac{\kappa_{ij}}{2}\|\mX_i-\mP_{ij}^{(k)}\|_2^2+\lambda_1\sum_i\left\|\Phi(\mX_i)\right\|_1\\
%     &+\sum_i\mB_i(\mX_i-\mE_i)+\frac{\rho}{2}\|\mX_i-\mE_i\|_2^2\,.
% \end{split}
% \end{align}
% Rearranging the terms in \eqref{eq:kp1_opt_xi}, it is equivalent to the following
\begin{align}
\label{eq:kp1_opt_xi_1}
    \min_{\mX_i}\quad \sum_i\frac{1}{2}\left\|\mX_i-\mV_i^{(k)}\right\|^2+\frac{\lambda_1}{\rho+\sum_j\kappa_{ij}}\|\Phi(\mX_i)\|_1\,,
\end{align}
where $\mV_i^{(k)}$ is 
\begin{align}
    \mV_i^{(k)} = \frac{\sum_j\kappa_{ij}\mQ_{ij}^{(k)}-\mB_i+\rho\mE_i}{\rho+\sum_j\kappa_{ij}}\,.
\end{align}
%The $l_1$-minimization problem in \eqref{eq:kp1_opt_xi_1} can be easily solved using soft-thresholding operator.
It can be easily solved using FISTA \cite{Beck:FISTA:2009}.
% \begin{align}
%     \mX_i=\Phi^{-1}\left(\Gamma\left(\Phi(\mV_i^{(k)}), \frac{\lambda_1}{\rho+\sum_j\kappa_{ij}}\right)\right)\,,
% \end{align}
% where $\Phi^{-1}(\cdot)$ is the inverse wavelet transform operator, $\Gamma(\cdot)$ is the soft-thresholding operator
% \begin{align}
%     \Gamma(x,\tau)=\left\{
%     \begin{array}{l}
%         (|x|-\tau)\cdot\textrm{sign}(x)  \\
%         0 
%     \end{array}
%     \quad
%     \begin{array}{l}
%         \textrm{if }|x|>\tau  \\
%         \textrm{if }|x|\leq\tau\,. 
%     \end{array}
%     \right.
% \end{align}

\subsection{Computation of $H_0$ and $R_2^*$}
We use proximal gradient regularization on the following $\mathscr{q}_i(\mH_0,\mR_2^*)$ with respect to $\mH_0$ and $\mR_2^*$ to solve \eqref{eq:admm_h_r}.
\begin{align}
    \mathscr{q}_i(\mH_0,\mR_2^*)=\left\|\mE_i\mH_0-t_i\mE_i\mR_2^*-\mE_i\log\mE_i\right\|_2^2
\end{align}
\subsubsection{Computation of $H_0$}
In the $(k+1)$-th iteration, we have
\begin{align}
\begin{split}
    % \mathscr{q}_i(\mH_0)\leq& \mathscr{q}_i(\mH_0^{(k)})+\langle \mH_0-\mH_0^{(k)},\nabla \mathscr{q}_i(\mH_0^{(k)}) \rangle\\
    % &+\frac{\gamma_i}{2}\|\mH_0-\mH_0^{(k)}\|_2^2\\
    % =&\frac{\gamma_i}{2}\|\mH_0-\mF_i^{(k)}\|_2^2+\mathcal{O}(\mH_0^{(k)})\,,
    \mathscr{q}_i(\mH_0)\leq\frac{\gamma_i}{2}\|\mH_0-\mF_i^{(k)}\|_2^2+\mathcal{O}(\mH_0^{(k)})\,,
\end{split}
\end{align}
where $\gamma_i$ is the Lipschitz constant of $\mathscr{q}_i(\mH_0)$ to ensure the proximal regularization holds, $\mathcal{O}(\mH_0^{(k)})$ is some constant the depends on the previous solution $\mH_0^{(k)}$, and $\mF_i^{(k)}$ is
\begin{align}
%    \nabla \mathscr{q}_i(\mH_0)&=2\mE_i^2(\mH_0-t_i\mR_2^*-\mE_i\log\mE_i)\\
    \mF_i^{(k)}&=\mH_0^{(k)}-\frac{1}{\gamma_i}\nabla \mathscr{q}_i(\mH_0^{(k)})\,.
\end{align}
We then solve the following $l_1$-minimization problem
% \begin{align}
%     \min_{\mH_0}\quad\sum_i\frac{\gamma_i}{2}\|\mH_0-\mF_i^{(k)}\|_2^2+\lambda_2\|\Phi(\mH_0)\|_1\,.
% \end{align}
% It is equivalent to
\begin{align}
    \min_{\mH_0}\quad\frac{1}{2}\left\|\mH_0-\frac{\sum_i\gamma_i\mF_i^{(k)}}{\sum_i\gamma_i}\right\|_2^2 + \frac{\lambda_2}{\sum_i\gamma_i}\|\Phi(\mH_0)\|_1\,.
\end{align}
% We can compute $\mH_0$ using the soft-thresholding as follows
% \begin{align}
%     \mH_0 = \Phi^{-1}\left(\Gamma\left(\frac{\sum_i\gamma_i\mF_i^{(k)}}{\sum_i\gamma_i},\frac{\lambda_2}{\sum_i\gamma_i}\right)\right)\,.
% \end{align}

\subsubsection{Computation of $R_2^*$}
The relaxation rate $\mR_2^*$ can be computed similarly. In the $(k+1)$-th iteration, we have
\begin{align}
\begin{split}
    % \mathscr{q}_i(\mR_2^*)\leq& \mathscr{q}_i({\mR_2^*}^{(k)})+\langle \mR_2^*-{\mR_2^*}^{(k)},\nabla \mathscr{q}_i({\mR_2^*}^{(k)}) \rangle\\
    % &+\frac{\nu_i}{2}\|\mR_2^*-{\mR_2^*}^{(k)}\|_2^2\\
    % =&\frac{\nu_i}{2}\|\mR_2^*-\mG_i^{(k)}\|_2^2+\mathcal{O}({\mR_2^*}^{(k)})\,,
    \mathscr{q}_i(\mR_2^*)\leq\frac{\nu_i}{2}\|\mR_2^*-\mG_i^{(k)}\|_2^2+\mathcal{O}({\mR_2^*}^{(k)})\,,
\end{split}
\end{align}
where $\nu_i$ is the Lipschitz constant of $\mathscr{q}_i(\mR_2^*)$ to ensure the proximal regularization holds, $\mathcal{O}({\mR_2^*}^{(k)})$ is some constant that depends on the previous solution ${\mR_2^*}^{(k)}$, and $\mG_i^{(k)}$ is
\begin{align}
%    \nabla \mathscr{q}_i(\mR_2^*)&=2t_i\mE_i^2(t_i\mR_2^*+\mE_i\log\mE_i-\mH_0)\\
    \mG_i^{(k)}&={\mR_2^*}^{(k)}-\frac{1}{\nu_i}\nabla \mathscr{q}_i({\mR_2^*}^{(k)})\,.
\end{align}
We then solve the following $l_1$-minimization problem
% \begin{align}
%     \min_{\mR_2^*}\quad\sum_i\frac{\nu_i}{2}\|\mR_2^*-\mG_i^{(k)}\|_2^2+\lambda_3\|\Phi(\mR_2^*)\|_1\,.
% \end{align}
% It is equivalent to
\begin{align}
    \min_{\mR_2^*}\quad\frac{1}{2}\left\|\mR_2^*-\frac{\sum_i\nu_i\mG_i^{(k)}}{\sum_i\nu_i}\right\|_2^2 + \frac{\lambda_3}{\sum_i\nu_i}\|\Phi(\mR_2^*)\|_1\,.
\end{align}
% We can compute $\mR_2^*$ using the soft-thresholding as follows
% \begin{align}
%     \mR_2^* = \Phi^{-1}\left(\Gamma\left(\frac{\sum_i\nu_i\mG_i^{(k)}}{\sum_i\nu_i},\frac{\lambda_3}{\sum_i\nu_i}\right)\right)\,.
% \end{align}

\subsection{Computation of $E_i$}
Let $\mD_i=\log\mE_i$ and $\mW_i=\mH_0-t_i\mR_2^*$. Here we assume $\mE_i\in [e_{\min}, e_{\max}]$ and $e_{\min}>0$ is close to zero so that $\mD_i$ is bounded. In practice we can choose $e_{\min}$ to be the machine precision. The optimization problem in \eqref{eq:admm_e} is equivalent to
\begin{align}
\label{eq:admm_d}
\begin{split}
    \min_{\mD_i}\quad \mathscr{l}(\mD_i)=&\frac{\rho}{2}\|\mX_i-\exp(\mD_i)\|_2^2+\lambda\exp(2\mD_i)\|\mD_i-\mW_i\|_2^2\\
    &+B_i(\mX_i-\exp(\mD_i))\,,
\end{split}
\end{align}
where $\mD_i\in[d_{\min},d_{\max}]$ is bounded. Although the above \eqref{eq:admm_b} is nonconvex, it consists of simple one-dimensional (pixel-wise) nonconvex problems. The \emph{global} minimizing solutions occur at either the boundaries or the stationary points that make the first order derivative $\mathscr{l}^\prime(d_i)=0$.

Since $\mD_i$ is bounded, we next show how to find the stationary points using the bisection method. We have
\begin{align}
\mathscr{l}^\prime(d_i)=&\exp(d_i)\cdot \mathscr{l}_1(d_i)\\
\begin{split}
    \mathscr{l}_1(d_i)=&\rho\big(\exp(d_i)-x_i\big)+2\lambda\exp(d_i)\cdot(d_i-w_i)\\
    &+2\lambda\exp(d_i)\cdot(d_i-w_i)^2-y_i\,.
\end{split}
\end{align}
%We can see that $d_i=-\infty$ is one stationary point. 
In order to use the bisection method to find the other stationary points that make $\mathscr{l}_1(d_i)=0$, we need to find intervals where $\mathscr{l}_1(d_i)$ is monotonically increasing or decreasing. We further compute the first order derivative of $\mathscr{l}_1(d_i)$ as follows
\begin{align}
\mathscr{l}_1^\prime(d_i)=&\exp(d_i)\cdot \mathscr{l}_2(d_i)\\
\begin{split}
    \mathscr{l}_2(d_i)=&2\lambda\cdot d_i^2+(6\lambda-4\lambda w_i)\cdot d_i \\
    &+\rho-6\lambda w_i+2\lambda+2\lambda w_i^2\,.
\end{split}
\end{align}
The above $\mathscr{l}_2(d_i)$ is a second-degree polynomial. We can find the monotonic intervals of $\mathscr{l}_1(d_i)$ based on the roots of $\mathscr{l}_2(d_i)$:
\begin{enumerate}
    \item If $\mathscr{l}_2(d_i)$ has less than two real roots, $\mathscr{l}_1^\prime(d_i)\geq 0$. We have that $[d_{\min},d_{\max}]$ is a monotonic interval of $\mathscr{l}_1(d_i)$.
    \item If $\mathscr{l}_2(d_i)$ has two real roots $d_i(1)<d_i(2)$. We have that
    \begin{itemize}
        \item $\mathscr{l}_1(d_i)$ is increasing in $(-\infty, d_i(1)]\bigcap[d_{\min},d_{\max}]$.
        \item $\mathscr{l}_1(d_i)$ is decreasing in $(d_i(1), d_i(2)]\bigcap[d_{\min},d_{\max}]$.
        \item $\mathscr{l}_1(d_i)$ is increasing in $(-d_i(2), \infty)\bigcap[d_{\min},d_{\max}]$.
    \end{itemize}
\end{enumerate}

Computing the stationary points is just the first step in finding the global minimum in \eqref{eq:admm_d}. They could be local minimum, local maximum or a saddle point. In order to find the global minimum, we still need to compare the function values of $q(d_i)$ at the stationary points with those at the boundaries of the monotonic intervals.
% For a particular monotonic interval $[a,b]$, we have
% \begin{enumerate}
%     \item If $\mathscr{l}_1(a)\mathscr{l}_1(b)>0$, no stationary point can be found in $[a,b]$. We need to compare the values of $q(a),q(b)$ and choose the one with smaller value as the minimizing solution in $[a,b]$.
%     \item If $\mathscr{l}_1(a)\mathscr{l}_1(b)\leq0$, a stationary point $d_s$ can be found in $[a,b]$. We need to compare the values of $q(d_s),q(a),q(b)$ and choose the one with the smallest value as the minimizing solution in $[a,b]$.
% \end{enumerate}
Eventually, by comparing the function values of all the minimizing solutions from every monotonic interval, we can find the global minimum in \eqref{eq:admm_d}.

\subsection{Convergence Analysis}
\label{subsec:convergence_admm}
ADMM has been used with success in nonconvex problems such as matrix completion \cite{Xu:ADMC:2012,Shen:ALMC:2014}, phase retrieval \cite{Wen:ADMM_PR:2012}, image denoising \cite{Lai:ADMM_Image:2014,Chan:ADMM_PP:2017}, etc. Characterizing its convergence behavior under the nonconvex setting has attracted a lot of interests in recent years \cite{Wang:ADMMNonconvex:2015,Li:ADMM_nonconvex:2015,Hong:ADMM_nonconvex:2016}. By examining the sufficient convergence conditions introduced in \cite{Wang:ADMMNonconvex:2015}, we can analyze the behavior of ADMM for $T_2^*$ mapping. We should note that even if the conditions are not satisfied, it does not mean that ADMM would diverge. It is still an open problem to establish necessary and sufficient conditions for ADMM to converge under the nonconvex setting. Nonetheless, the following analysis still provides valuable insights into the factors that influence the behavior of the proposed approach.

There are two sets of variables in the constrained nonconvex problem \eqref{eq:t2_star_mapping_separate}: $\mathcal{V}_1=\{\mZ_i,\mX_i\}$ and $\mathcal{V}_2=\{\mE_i,\mH_0,\mR_2^*\}$. In the following analysis we shall assume the feasible set $\{\mathcal{V}_1,\mathcal{V}_2\}$ is bounded. Letting $\Psi(\mathcal{V}_1,\mathcal{V}_2)=f(\mathcal{V}_1)+\lambda\cdot g(\mathcal{V}_2)$, we can rewrite \eqref{eq:t2_star_mapping_separate} as follows
\begin{align}
    \label{eq:t2_star_mapping_separate_rewrite}
    \begin{split}
        \min_{\mathcal{V}_1,\mathcal{V}_2}&\quad\Psi(\mathcal{V}_1,\mathcal{V}_2)\\
        \textrm{subject to}&\quad\mT_1\mathcal{V}_1+\mT_2\mathcal{V}_2=0\,,
    \end{split}
\end{align}
where $\mT_1,\mT_2$ are matrices used to create the constraint $\mX_i=\mE_i$. We next examine the sufficient convergence conditions outlined in \cite{Wang:ADMMNonconvex:2015} one by one:
% \begin{align}
% \label{eq:t2_star_mapping_separate}
% \begin{split}
%     \min_{\mZ_i,\mX_i,\mH_0,\mR_2^*,\mE_i}&\quad f(\mZ_i,\mX_i)+\lambda\cdot g(\mE_i,\mH_0,\mR_2^*)\\
%     \textrm{subject to}&\quad \mX_i=\mE_i\,.
% \end{split} \tag{\eqref{eq:t2_star_mapping_separate} revisited}
% \end{align}

\begin{enumerate}[leftmargin=*]
    \item The objective function $\Psi(\mathcal{V}_1,\mathcal{V}_2)$ is coercive, that is, $\Psi\rightarrow\infty$ when $\|(\mathcal{V}_1,\mathcal{V}_2)\|\rightarrow\infty$.
    \item The images (Im) of $\mT_1,\mT_2$ are the same, $\mathrm{Im}(\mT_1)=\mathrm{Im}(\mT_2)=\mathrm{Im}([\mI\  \vzero])$, where $\mI$ is the identity matrix and $\vzero$ is an all zero vector.
    \item With $\{\mZ_i,\mX_i\}$ fixed and $\mE_i=\mX_i$, we have the following problem to solve
    \begin{align}
    \label{eq:h_r}
    \begin{split}
        \min_{\mH_0,\mR_2^*}\quad &\textstyle\sum_i\mX_i^2\|\mH_0-t_i\mR_2^*-\log\mX_i\|_2^2\\
        &+\lambda_2\|\Phi(\mH_0)\|_1+\lambda_3\|\Phi(\mR_2^*)\|_2\,.
    \end{split}
    \end{align}
    We need at least two echo images for \eqref{eq:h_r} to have a unique solution. Since the feasible set $\{\mH_0,\mR_2^*\}$ is bounded, the solution is Lipschitz continuous with respect to the input $\mX_i$.
    \item With $\{\mH_0,\mR_2^*,\mE_i\}$ and $\mX_i$ fixed, we have the following problem to solve
    \begin{align}
    \label{eq:z}
        \min_{\mZ_i}\quad\textstyle\sum_{j}\|\mY_{ij}-\mA_{ij}\mX_i\mZ_i\|_2^2\,,
    \end{align}
    Sufficient incoherent measurements are needed for \eqref{eq:z} to have a unique solution. Although we are undersampling in the k-space, additional incoherent measurements can be acquired by the multiple receiver coils. Since the feasible set of $\mZ_i$ is bounded, the solution is also Lipschitz continuous with respect to the input $\mX_i$.
    \item With $\{\mH_0,\mR_2^*,\mE_i\}$ and $\mZ_i$ fixed, we solve the following problem subject to $\mX_i=\mE_i$
    \begin{align}
        \label{eq:xi}
        \min_{\mX_i}&\quad\textstyle\sum_j\|\mY_{ij}-\mA_{ij}\mZ_i\mX_i\|_2^2+\lambda_1\|\Phi(\mX_i)\|_1\,.
        %\textrm{subject to}&\quad\mX_i=\mE_i\,.
    \end{align}
    The above \eqref{eq:xi} simply has a unique solution $\mX_i=\mE_i$ and the solution is Lipschitz continuous with respect to the input $\mE_i$.
\end{enumerate}
In summary, apart from the assumption that the feasible set $\{\mathcal{V}_1,\mathcal{V}_2\}$ is bounded, we need sufficient incoherent measurements by multiple receiver coils from at least two echo times so that sufficient convergence conditions would hold.

\subsection{Parameter Tuning}
The regularization parameters $\{\lambda_1,\lambda_2,\lambda_3,\lambda,\rho\}$ needs to be properly tuned in order to achieve best performance. With the five parameters at hand, it would be computationally inefficient to tune them all at once. Here we take the divide-and-conquer strategy and tune them in different groups. In general, the optimal parameters $\{\lambda_1,\lambda_2,\lambda_3\}$ obtained for the decoupled recovery in \eqref{eq:decoupled_xi}-\eqref{eq:decoupled_h0_r2} can be migrated to the joint recovery. We can then fix $\{\lambda_1,\lambda_2,\lambda_3\}$ and focus on tuning $\{\lambda,\rho\}$ in the joint recovery. In summary, we can tune the parameters in the following order:
\begin{enumerate}
    \item Find the optimal $\lambda_1$ in \eqref{eq:decoupled_xi} that recovers $\mZ_i,\mX_i$.
    \item Find the optimal $\lambda_2,\lambda_3$ in \eqref{eq:decoupled_h0_r2} that recover $\mH_0,\mR_2^*$.
    \item Fix the $\lambda_1,\lambda_2,\lambda_2$ obtained previously, find the optimal $\lambda,\rho$ in \eqref{eq:t2_star_mapping} that solve the joint recovery problem.
\end{enumerate}
From the \emph{in vivo} experiments we learned that the optimal parameters are \emph{stable} and \emph{generalizable} on 3D MRI data that are acquired under the same protocol. This greatly simplifies the overall parameter tuning process. In practice we can tune the parameters on a fully sampled 2D slice, aka training data, and use them to reconstruct other undersampled 2D slices in a 3D volume.

\section{Experimental Results}
\label{sec:exp}
We acquired \emph{in vivo} 3D brain data on a 3T MRI scanner (Siemens Prisma), with written consent obtained from the subject before imaging and approval from the Institutional Review Board of Emory University. The k-space was fully sampled during the acquisition to provide the ``gold standard'' references for evaluation. The undersampling took place in the phase encoding $y-z$ plane afterwards according to the randomly generated Poisson disk sampling patterns. The readout direction $x$ was always fully sampled at each TE. The GRE sequence was used for $T_2^*$ mapping, and the data were acquired with a 32-channel head coil. Two acquisition protocols are adopted to acquire data from 6 subjects. 
\begin{itemize}
\item For the first protocol, we have the number of echoes = 6, the first echo time = 7.64 ms, echo spacing = 5.41 ms, slice thickness = 0.7 mm, resolution = 0.6875 mm, pixel bandwidth = 260 Hz, TR = 40 ms, and FoV = 22 cm, which takes approximately 35 minutes to finish. 
\item For the second protocol, we have the number of echoes = 4, the first echo time = 7.32 ms, echo spacing = 8.68 ms, slice thickness = 0.7 mm, resolution = 0.6875 mm, pixel bandwidth = 260 Hz, TR = 38 ms, and FoV = 22 cm, which takes approximately 33 minutes to finish.
\end{itemize}
We eventually saved the fully sampled 3D k-space data in a matrix of size $320\times320\times280$ for each subject.

As discussed in section \ref{sec:problem_formulation}, the 3D reconstruction problem can be decomposed into parallelizable 2D problems to speed up the reconstruction process. In the following we shall compare different reconstruction approaches and sampling schemes on uniformly selected 2D slices that cover the region of interest. Taking the recovered $\hat{\mR}_2^*$ image for example, we compute the pixel-wise relative errors of the brain\footnote{A mask is used to extract the subject's brain.} $\frac{|r_2^*-\hat{r}_2^*|}{|r_2^*|}$ according to the gold standard $\mR_2^*$ reconstructed from fully-sampled data, and use the average relative error of all brain pixels in the 2D slice as the comparison criterion. 

% \begin{figure}[tbp]
% \centering
% \subfigure{
% \label{fig:r2_rel_error_compare}
% \includegraphics[width=0.23\textwidth]{figures/R2_rel_error_compare_ave_all_data.pdf}}
% \subfigure{
% \label{fig:x0_rel_error_compare}
% \includegraphics[width=0.23\textwidth]{figures/X0_rel_error_compare_ave_all_data.pdf}}
% \caption{Comparison of the proposed decoupled and joint approaches with the state-of-the-art direct approach across different sampling rates. The average relative error of each approach is computed using reconstructions from 6 different subjects.}
% \label{fig:rel_error_compare}
% \end{figure}

\begin{figure*}[tbp]
\centering
\subfigure{
\label{fig:r2_rel_error_compare}
\includegraphics[width=0.45\textwidth]{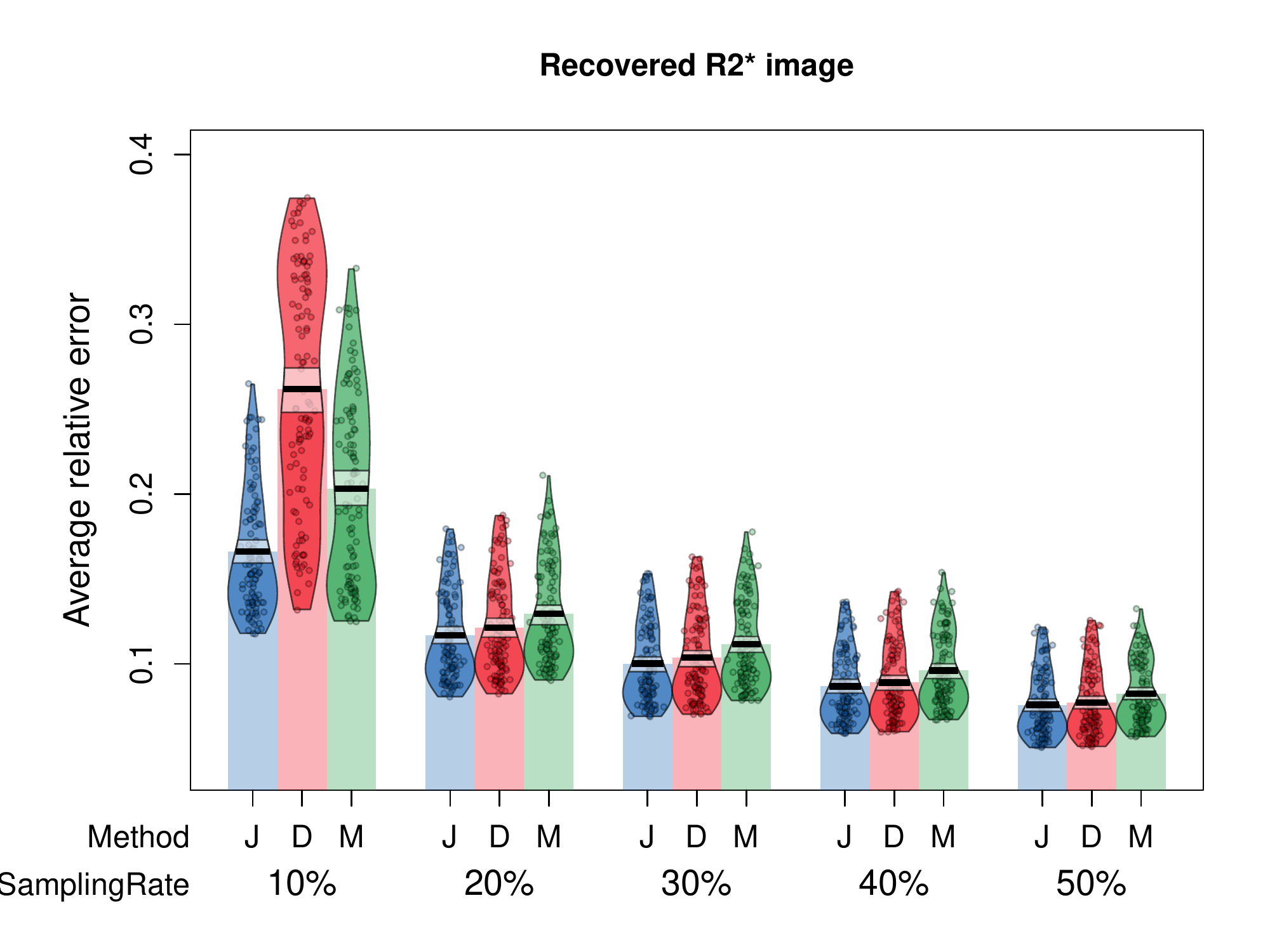}}
\subfigure{
\label{fig:x0_rel_error_compare}
\includegraphics[width=0.45\textwidth]{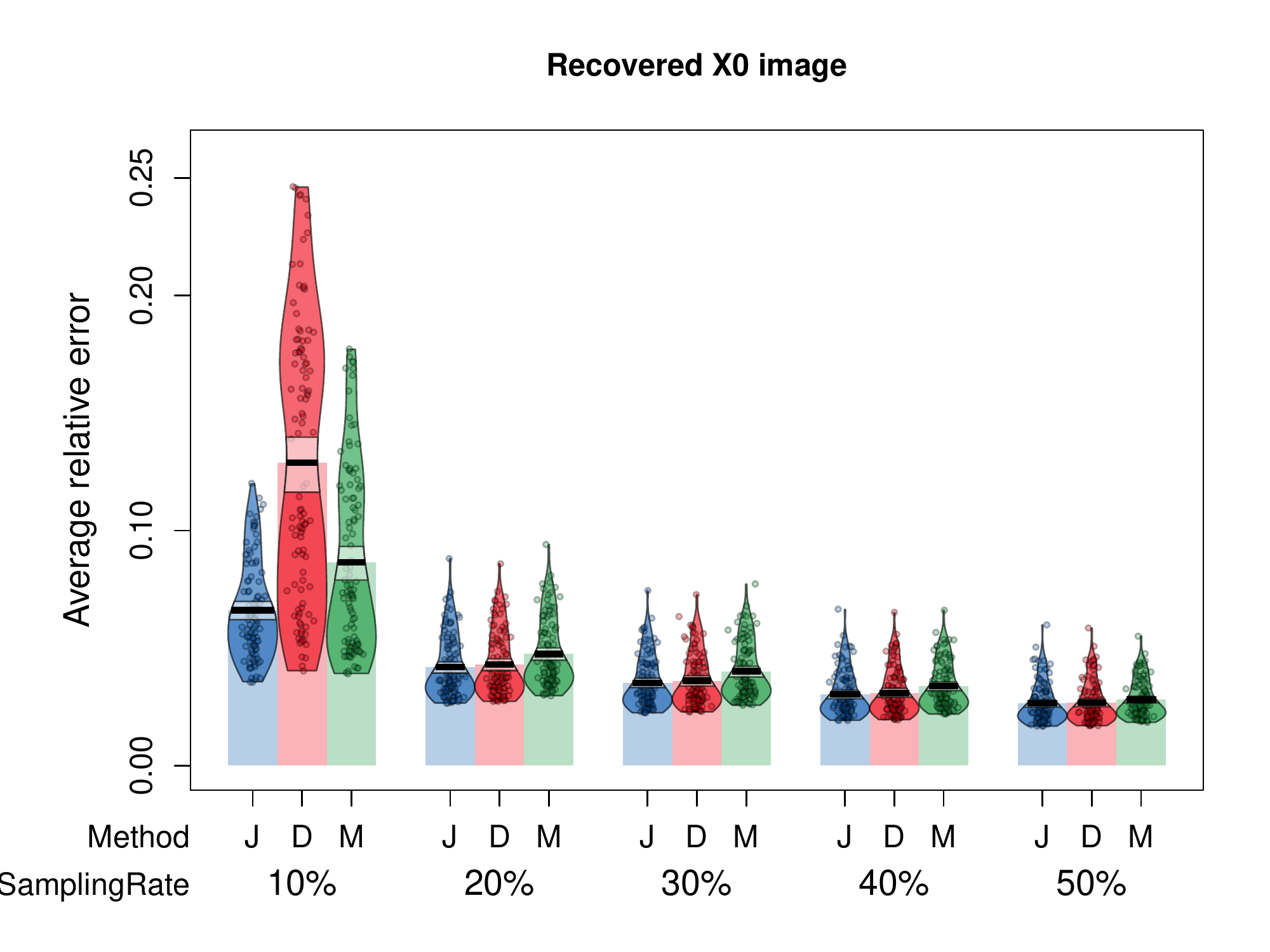}}
\caption{Comparison of the proposed joint (\textbf{J}) and decoupled (\textbf{D}) recovery approaches with the state-of-the-art model-based (\textbf{M}) approach across different sampling rates. Every point here corresponds to the average relative error of all brain pixels in one 2D slice.}
\label{fig:rel_error_compare}
\end{figure*}

\begin{figure*}[tbp]
\centering
\subfigure{
\label{fig:rel_error_compare_sd009_10}
\includegraphics[width=0.9\textwidth]{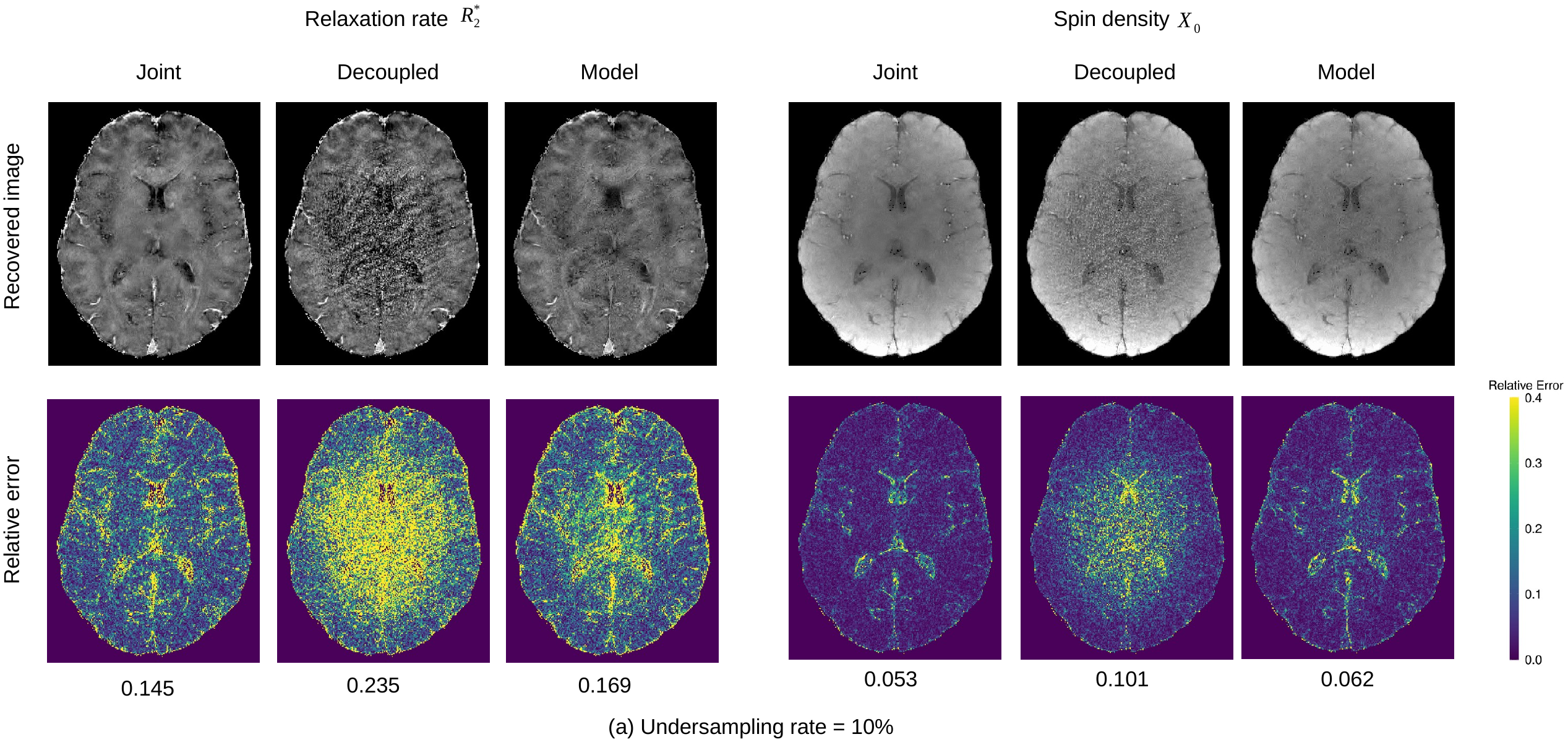}}
\subfigure{
\label{fig:rel_error_compare_sd009_10}
\includegraphics[width=0.9\textwidth,trim={0cm 0cm 0cm 0.6cm},clip]{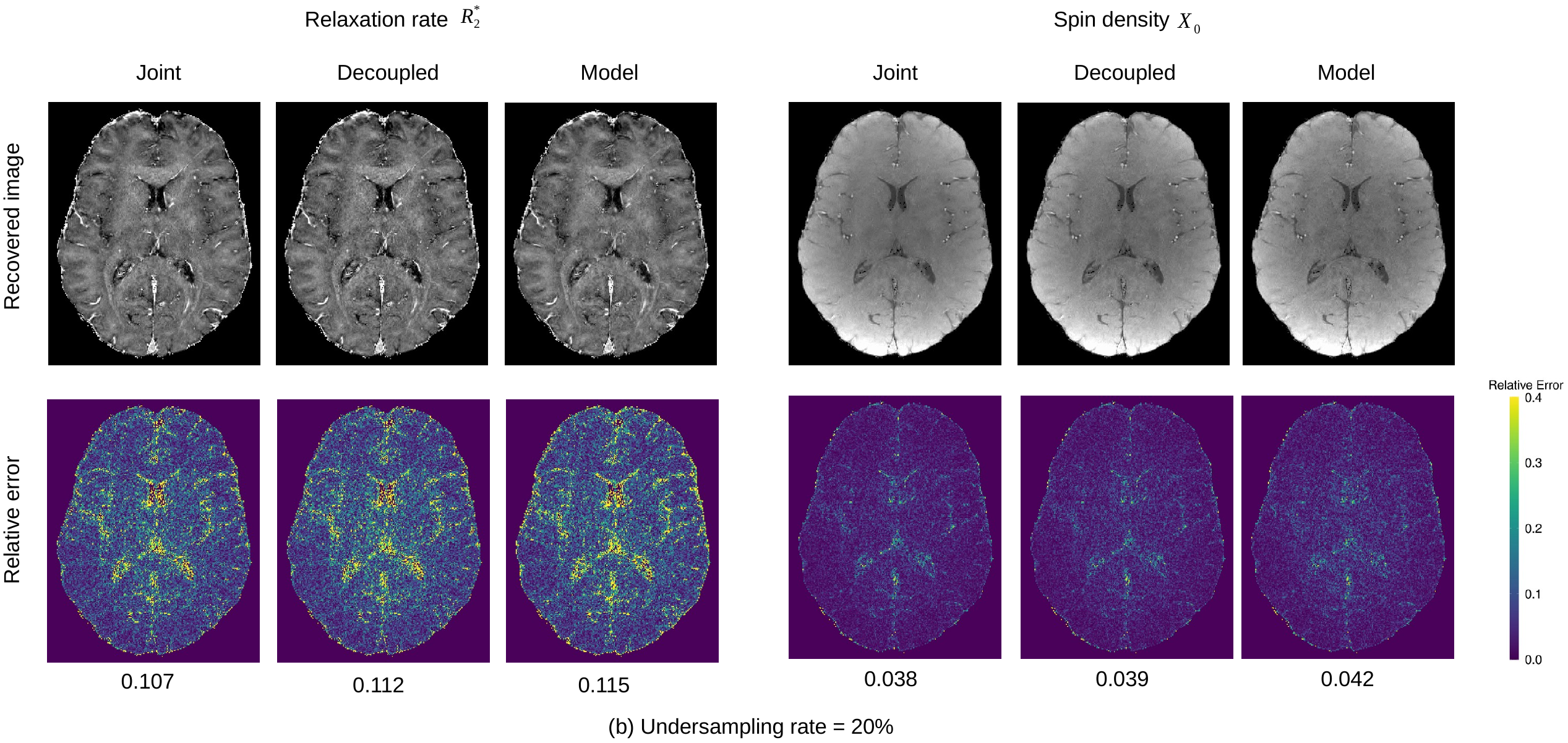}}
\caption{The recovered $\mR_2^*$ and $\mX_0$ images using the proposed joint and decoupled recovery approaches, and the state-of-the-art model-based approach when the undersampling rates are $10\%$ and $20\%$.}
\label{fig:rel_error_compare_sd009}
\end{figure*}

\begin{figure*}[tbp]
\centering
\vspace{-2em}
\subfigure{
\label{fig:r2_same_diff}
\includegraphics[width=0.3544\textwidth]{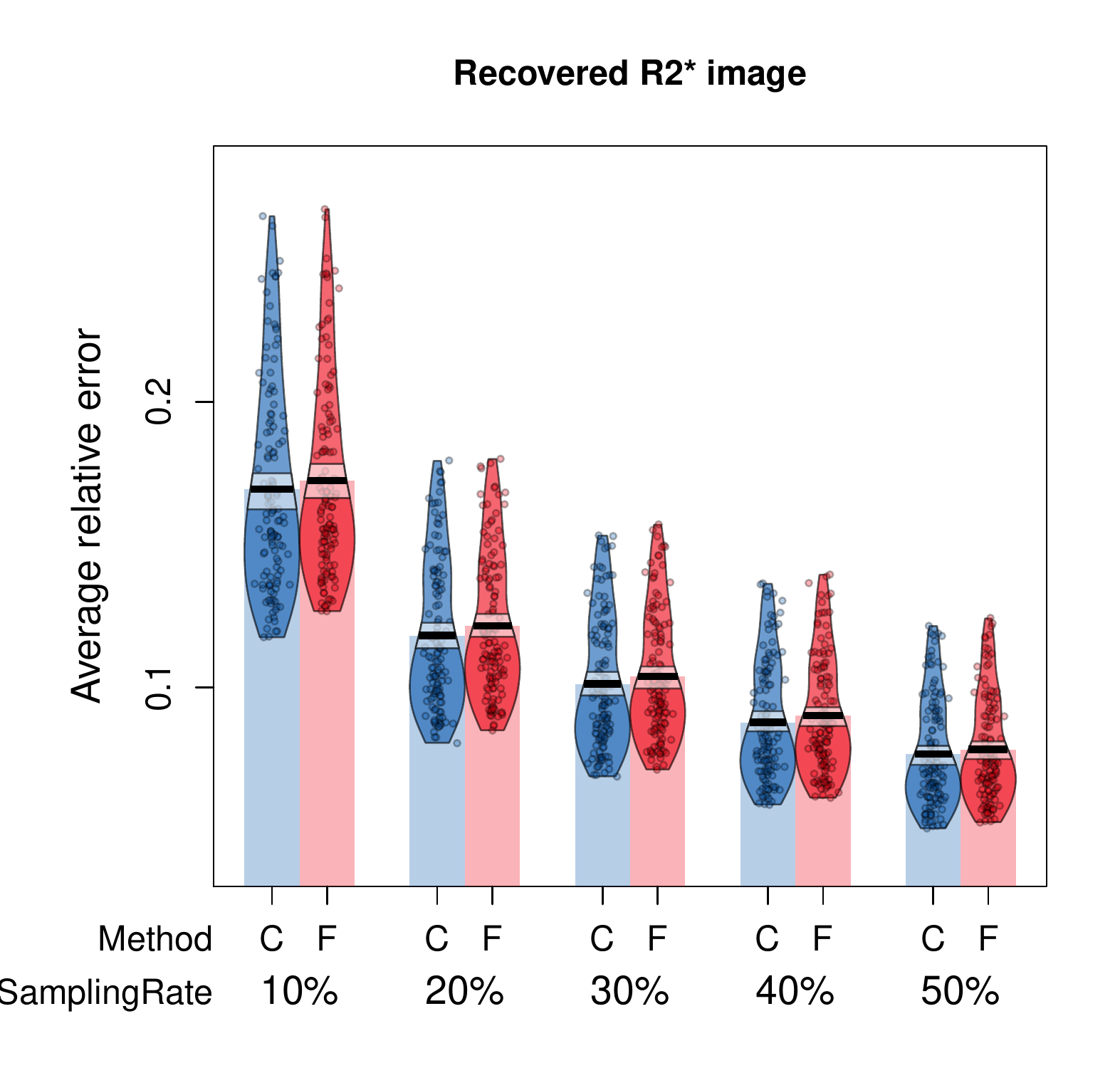}}
\subfigure{
\label{fig:x0_same_diff}
\includegraphics[width=0.3544\textwidth]{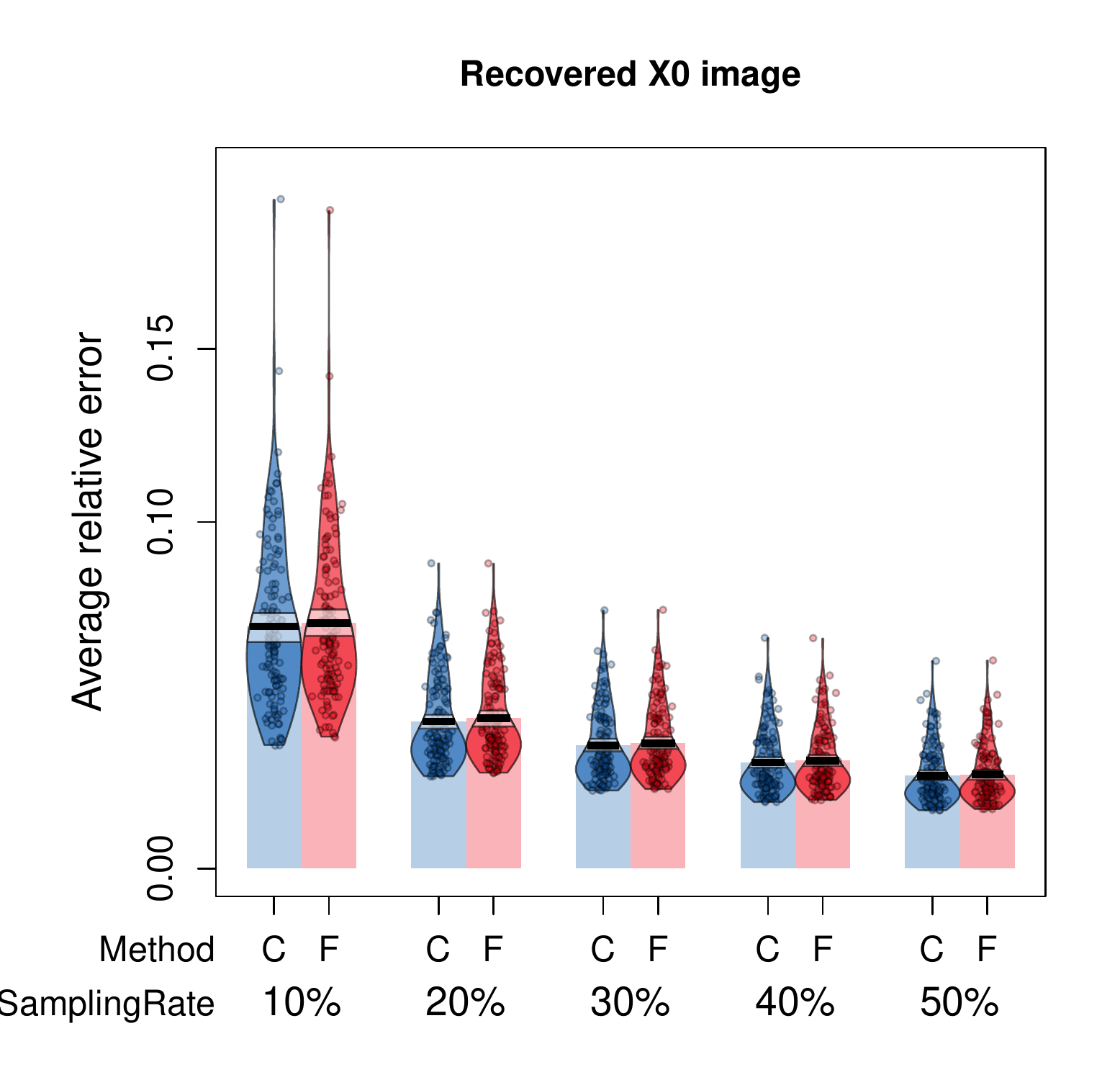}}
\vspace{-1em}
\caption{Comparison of the complimentary (\textbf{C}) and fixed (\textbf{F}) Poisson disk sampling patterns in the joint recovery approach across different sampling rates. Every point here corresponds to the average relative error of all brain pixels in one 2D slice.}
\vspace{-1em}
\label{fig:x0_r2_same_diff}
\end{figure*}

\subsection{Comparison of Reconstruction Approaches}
\label{subsec:exp_undersampling}
We compare the proposed decoupled and joint recovery approaches with the state-of-the-art model-based approach that solves the following problem \cite{Block:ModelT2:2009} 
\begin{align}
\label{eq:direct_x0_r2}
\begin{split}
    \min_{\mZ_i,\mX_0,\mR_2^*}\quad&\sum_{ij}\|\mY_{ij}-\mA_i\mS_j\mZ_i\mX_0\cdot\exp(-t_i\cdot\mR_2^*)\|_2^2\\
    &+\lambda_1\|\Phi(\mX_0)\|_1+\lambda_2\|\Phi(\mR_2^*)\|_1\,.
\end{split}
\end{align}
We set $d_{\min}=2$ in the Poisson disk sampling scheme and use different sampling patterns across the echo times. The parameters for each approach are individually tuned to achieve best performance. 

%\subsubsection{Reconstruction from Undersampled Measurements}
We compare the three approaches on the 3D MRI dataset acquired from 6 subjects. An acquisition protocol with 6 echoes is adopted for the first three subjects, and a different acquisition protocol with 4 echoes is adopted for the other three subjects. As long as the acquisition protocol is fixed, the optimal parameters tuned on one 2D slice of a subject can be used to perform reconstruction on the other subjects. We undersample the the k-space measurements with the sampling rates varying between $10\%$ and $50\%$. For each subject, we uniformly choose 20 out of 208 2D slices, and use the three approaches to recover the relaxation rate $\mR_2^*$ and the spin density $\mX_0$ from undersampled data. The average relative errors of all the 2D slices across 6 subjects are computed and shown in Fig. \ref{fig:rel_error_compare}. We can see that the joint recovery approach generally performs better than the other two approaches. When the sampling rate is low ($\sim10\%$), the joint recovery approach performs much better than the decoupled approach and the model-based approach. When the sampling rate is higher ($\geq20\%$), the decoupled and joint recovery approaches perform almost equally well, with a very mild advantage from the joint recovery approach. As the sampling rate increases towards $50\%$, the performances of the three approaches become similar.

We next use one of the 2D slices as an example, and show the recovered $\hat{\mR}_2^*$ and $\hat{\mX}_0$ images in Fig. \ref{fig:rel_error_compare_sd009}. We can see that the joint recovery approach does a much better job in reconstructing the central brain region in the low-sampling rate regime ($10\%$). The joint recovery approach not only enforces the monoexponential decay model during the reconstruction, but also imposes the multi-echo regularization of echo images $\mX_i$. This allows it to take in more prior information to help with the reconstruction from undersampled measurements. As a comparison, the decoupled approach does not incorporate the monoexponential decay model in the reconstruction of the echo images $\mX_i$, while the model-based approach in \eqref{eq:direct_x0_r2} does not impose multi-echo regularization on the echo images $\mX_i$. This puts the other two approaches in a disadvantageous position, especially when the data is highly undersampled.

\subsection{Comparison of Sampling Schemes}
The Poisson disk sampling scheme enforces a minimum distance $d_{\min}$ between any two sampling locations. The choice of $d_{\min}$ has a direct effect on the reconstruction performance, and we choose $d_{\min}=2$ pixels in the experiments. The undersampling process at the echo times are independent, which gives us the freedom to choose different sampling patterns at different echo times. Using the joint recovery approach for reconstruction, we next compare the case where the sampling patterns are fixed and the case where the sampling patterns are complementary to one another across different echoes. The experimental settings are kept the same as in section \ref{subsec:exp_undersampling}. The average relative errors of the recovered $\hat{\mR}_2^*$ and $\hat{\mX}_0$ images are shown in Fig. \ref{fig:x0_r2_same_diff}. We can see that the two types of sampling patterns perform almost equally well. The complementary sampling patterns perform only slightly better. Based on this experiment, we can keep the sampling pattern fixed across different echoes without sacrificing the reconstruction performance too much, thus simplifying the pulse sequence programming on the MRI scanner.

\section{Conclusion and Discussion}
\label{sec:conclusion}
In this paper we aim to develop fast 3D $T_2^*$ imaging method that reconstructs the relaxation rate $\mR_2^*=\frac{1}{\mT_2^*}$ and the spin density $\mX_0$ from undersampled measurements in quantitative MRI. We formulate the reconstruction problem into two subproblems: one that recovers the multi-echo images $\mX_i$, and one that recovers $\mR_2^*,\mX_0$. They can be solved separately via the standard approach or jointly the ADMM. Compared to previous approaches that only enforce sparse priors on $\mX_0$ and $\mR_2^*$, the propose approach makes use of additional sparse priors on the multi-echo images $\mX_i$ during the reconstruction. To avoid the scaling issue caused by the nonlinearity of the monoexponential decay model, we further derive its linear approximation to compute the regularized least square fit of $\mX_0$ and $\mR_2^*$. Experimental results show that the proposed joint recovery approach generally outperforms the state-of-the-art model-based approach, especially in the low-sampling rate regime.

The reconstruction of $\mR_2^*$ and $\mX_0$ is inherently a nonconvex problem. With the linear approximation of the monoexponential decay model, the decoupled approach in \eqref{eq:decoupled_xi}-\eqref{eq:decoupled_h0_r2} becomes convex and is easy to solve. However, it performs much worse than the nonconvex joint recovery approach in the low-sampling rate regime. We showed in section 
\ref{subsec:convergence_admm} that the ADMM used in joint recovery could still achieve convergence in the nonconvex setting when sufficient incoherent measurements are sampled from at least two echo times. When the sampling rate is high, the decoupled and joint approaches perform almost equally well. In this case, we can simply choose the decoupled approach for reconstruction.

\bibliographystyle{IEEEbib}
\bibliography{refs}

\end{document}